\documentclass[9pt,twocolumn,twoside]{osajnl}

\journal{josab} 

\setboolean{shortarticle}{false}

\title{Fundamental residual amplitude modulation in electro-optic modulators}

\author[1,*]{Alfredo E. Dom\'inguez }
\author[1]{Walter E. Ortega Larcher}
\author[2]{Carlos N. Kozameh}

\affil[1]{Facultad de Ingenier\'ia, Instituto Universitario Aeron\'autico, Centro Regional Universitario C\'ordoba,  C\'ordoba, Argentina}
\affil[2]{Instituto de F\'isica Enrique Gaviola, FaMAF, Universidad Nacional de C\'ordoba, C\'ordoba, Argentina}

\affil[*]{Corresponding author: adominguez@iua.edu.ar}

\begin{abstract}
The residual amplitude modulation ($\mathrm{RAM}$) is the undesired, non-zero amplitude modulation that usually
occurs when a phase modulation based on the electro-optic effect is imprinted on a laser beam. 
In this work, we show that electro-optic modulators (EOMs) that are used to generate the sidebands on the laser beam also generate a $\mathrm{RAM}$ in the optical setup. This result contradicts standard textbooks, which assume the amplitude remains unchanged in the process and should be considered as a fundamental $
\mathrm{RAM}$  ($\mathrm{RAM_{F}}$) for these devices. 
We present a classical model for the propagation of an infrared laser with frequency $\omega_{0}$ in a wedge-shaped crystal and an  EOM with an RF modulating signal of frequency $\Omega$. Since ${\Omega}\ll \omega_{0}$, we solve Maxwell's equations in a time-varying media via a WKB approximation and we write the electromagnetic fields in terms of quasi-plane waves.  From the emerging fields of the setup,   we compute the associated $\mathrm{RAM_{F}}$ and show that it 
depends on 
the phase-modulation depth  $m$ and the quotient $\left(\frac{\Omega}{\omega_{0}}\right)$. 
The $\mathrm{RAM_{F}}$ values obtained for the EOMs used in gravitational wave detectors are presented.  
Finally, the cancellation of $\mathrm{RAM_{F}}$ is analyzed.
\end{abstract}

\setboolean{displaycopyright}{true}

\begin{document}

\maketitle

\section{Introduction}

\qquad The electro-optic modulators (EOMs) are devices designed to modulate a laser beam. Depending on the configuration adopted by the EOM, they can be used to change the polarization state, to modulate the phase or the amplitude of the laser \cite{YarivQE}. It is also possible to simultaneously modulate the amplitude and phase of the beam  \cite{Cusack2004}.

The EOMs have multiple applications, for example, frequency modulation spectroscopy \cite{Bjorklund1980, Bjorklund1983}, modulation transfer spectroscopy \cite{Camy1982, Shirley1982}, two-tone frequency modulation spectroscopy \cite{Janik1986, Cooper1987}, laser frequency stabilization and cavity length locking 
\cite{Drever1983, Black2001}. Specifically, the EOMs are used in the Laser Interferometer Gravitational-Wave Observatory (LIGO) as well as in VIRGO where they play an important role. {These observatories, which in the year 2015 have achieved the first direct observation of gravitational waves emitted by black hole coalescence, are capable of detecting perturbations of the space-time on the order of $10^{-19}$ m} \cite{LIGO2016_2, LIGO2017, LIGO2016}.

To achieve these sensitivity levels it is necessary to accurately control the length of the two Fabry-Perot cavities, (each cavity 4 km in length) so that they are always in optical resonance. The length control system is done via a variation of the Pound-Drever-Hall technique to generate sidebands in the laser beams that go to the cavities. The sidebands are generated by EOM that produces a phase modulation in the laser beam \cite{Fritschel2001}. To achieve phase modulation, the index of refraction of the crystal used in the EOM is modulated by a periodic, slowly varying external electric field. This external field is perpendicular to the direction of the laser wave and both are aligned with the principal axis of the crystal.

However, besides the required phase modulation, the experimental setup of the EOMs also produced an unwanted modulation in the amplitude of the transmitted wave. This residual amplitude modulation ($\mathrm{RAM}$) could have pronounced effects on the optomechanical response of the interferometer. In fact, there is evidence that $\mathrm{RAM}$ affects the calibration of the Fabry-Perot cavities \cite{Kokeyama2014}. 

$\mathrm{RAM}$ in the aLIGO setup was attributed to deficiencies in the phase modulation process. According to several authors, there are many sources that could contribute to  $\mathrm{RAM}$: 
\begin{enumerate}[]
\item Etalon effect caused by the multiple reflections on the crystal faces \cite{Whittaker1985, Wong1985}.
\item Misalignment between the incident beam and the principal axis of the crystal \cite{Wong1985}.
\item Piezoelectric response of the crystal to the modulating frequency \cite{Ishibashi2002}.
\item Deformation of the crystal with the local temperature \cite{Ishibashi2002}.
\item Non-uniformity of the modulating electric field \cite{Ishibashi2002}.
\item Photorefractive effect on the crystal  
\cite{Sathian_2012_Intensity,Sathian_2013_Dependence,Sathian_2013_Reducing}.
\end{enumerate}

In order to avoid $1$ and $2$, the crystal of the EOM is wedge-shaped so that the laser direction is no longer perpendicular to the opposite faces of the crystal. The aLIGO configuration uses a Rubidium Titanyl Phosphate (RTP) wedge-shaped crystal for each EOM \cite{LVC}.

Experimental studies performed in \cite{Wu} analyzed the incidence of $3$ and $4$ for aLIGO laser power and showed that it is not relevant, given the current layout of aLIGO.
 
Although $\mathrm{RAM}$  has been reduced to low levels (ranging from $10^{-5}$ to $10^{-6}$) it has not been possible to eliminate it from the setup.  

It is important to note that $\mathrm{RAM}$ is always present in all technological designs which include EOM. In fact, experimental studies of $\mathrm{RAM}$ have been reported in a number of technological applications (\cite{Wong1985}, \cite{Jaatinen2008, Li2012, Zhang2014, Li2014, Shen2015}).

In this work, we show that the same process that produces a phase modulation for the propagating electromagnetic wave in a time-dependent index of refraction medium will also give a modulating amplitude for the transmitted wave.

The theoretical model proposed in this work comes directly from Maxwell's equations in a medium with a time-dependent index of refraction. Thus, it is not any of the proposed sources of $\mathrm{RAM}$ listed above. Directly from the field equations, one shows that $\mathrm{RAM}$ is an inherent feature of the physics of the problem and it is thus unavoidable. The mathematical expression of the transmitted electric field obeys all the parametric dependence mentioned in \cite{Kokeyama2014} and should be regarded as a $\mathrm{RAM_{F}}$. It is worth mentioning that the present values of $\mathrm{RAM}$ for aLIGO are still two orders of magnitude above this predicted limit. Hence, there is still plenty of room for improvement until they reach the lower limit.

In Section (\ref{sec:laserpropagation}) we provide a simple model to describe light propagation inside a crystal with an optical modulator. In Section (\ref{sec:Model_EOM}) we obtain the transmitted wave for our model EOM and we present the principal result of this work, the expression for $\mathrm{RAM_{F}}$ for our model of EOM. We compute the $\mathrm{RAM_{F}}$ associated with the aLIGO setup.
In section (\ref{sec:micro_macro}) we analyze $\mathrm{RAM_{F}}$ from the conservation of energy point of view. 
In section (\ref{sec:cancellation_RAM}) we analyze the suppression of $\mathrm{RAM_{F}}$ to different applications of EOMs.
Finally, in section (\ref{sec:conclusions})  we summarize our results and analyze the physical validity of the approximations used to obtain this $\mathrm{RAM_{F}}$ for the aLIGO laser and other different applications of EOM lasers.

\section{Laser propagation in a media with a time-dependent index of refraction}
\label{sec:laserpropagation}
\qquad The propagation of electromagnetic waves inside a crystal with a lossless and time-dependent media can be modeled by Maxwell's equations,

\begin{equation}
\mathbf{\nabla \times E} + \frac{\partial \mathbf{B} }{\partial t} = 0 ,
\label{eq:Faraday}
\end{equation}
\begin{equation}
\mathbf{\nabla \times H} - \frac{\partial \mathbf{D} }{\partial t}  = 0 ,
\label{eq:Ampere-Maxwell}
\end{equation}
\begin{equation*}
\mathbf{\nabla \cdot B }= 0 ,
\end{equation*}
\begin{equation*}
\mathbf{\nabla \cdot D} = 0,
\end{equation*}
together with the constitutive relations:
\begin{equation}
\mathbf{B} =\mu_{0} \ \mathbf{H},
\label{eq:BH}
\end{equation}
\begin{equation}
\mathbf{D} = \varepsilon(t) \  \mathbf{E}. 
\label{eq:ED}
\end{equation}

The above relations show that the medium is non-magnetic and that the permittivity is time-dependent. Moreover, we assume that the electro-optical effect is linear.

We recall that the electro-optical effect is the change in the magnitude and direction of the refractive indices in the crystal due to the presence of an external electric field. This field has a periodical time dependence 
(usually with a single frequency) and we will denote it as a modulating electric field.

We thus write the dielectric permittivity as,

\begin{equation*}
\varepsilon (t)= \varepsilon_0 \ n^{2}(t),
\end{equation*}
where $\varepsilon_{0}$ is vacuum permittivity and $n(t)$ is time-dependent index of refraction.

Remark: to observe an electro-optical effect the modulated electric field must vary slowly with time, i.e., its frequency must be much smaller than that of the laser beam propagating inside this media.
Denoting by $\Omega$  the frequency of the modulating field,  by $\omega_{0}$ the frequency of the electromagnetic wave, we assume that,
\begin{equation*}
\frac{\Omega}{\omega_{0}} \ll 1 .
\end{equation*}

Introducing the vector potential $\mathbf{A}$  in the Coulomb gauge
\begin{equation*}
\mathbf{\nabla \cdot A}=0,
\end{equation*}
we write the electric and magnetic fields of the laser wave as

\begin{equation*}
\mathbf{E} = - \frac{\partial \mathbf{A}}{\partial t},
\end{equation*}

\begin{equation*}
\mathbf{B} = \mathbf{\nabla \times  A},
\end{equation*}
where we have set the scalar potential $\Phi = 0$ since we are solving the source-free Maxwell equations. Outside the crystal, the scalar potential vanishes and since it satisfies an elliptic equation, by uniqueness, it vanishes everywhere.

Thus, the Ampere-Maxwell equation is the only non-trivial equation to be solved. Assuming that the incident wave is polarized in the $"\mathrm{Z}"$ direction and propagates in the $"\mathrm{Y}"$ direction we write,

\begin{equation*}
\mathbf{E} = \mathrm{E_{z}}(y,t) \ \mathbf{\hat{e}_{z}},
\end{equation*}

\begin{equation*}
        \mathrm{D_{z}}\left( y ,t \right) = \varepsilon(t) \ \mathrm{E_{z}} \left( y ,t \right).
\end{equation*}

It follows from the above conditions that the potential vector satisfies,

\begin{equation}
\frac{\partial^2 \mathrm{A_{z}}}{\partial y^2} - \mu_{0} \  \frac{\partial}{\partial t}\left( \varepsilon \
\frac{\partial \mathrm{A_{z}}}{\partial t} \right)= 0.
\label{main}
\end{equation}

To solve (\ref{main}), we write the solution as the real part of,
\begin{equation*}
\mathrm{A_{z}} = A_{\mathrm{0Z}}(y,t) \ \mathrm{e}^{\mathrm{i}\phi(y,t)}.
\end{equation*}

This particular waveform will be useful when considering the case of interest to us, the so-called quasi-static solution where the phase $\phi(y,t)$ is a small deviation from the static case, and the amplitude $A_{\mathrm{0Z}}(y,t)$ a slowly varying function of space and time.
Replacing  in (\ref{main}), we obtain two differential equations, one for the real part:

\begin{equation*}
\begin{aligned}
\frac{\partial^{2}A_{\mathrm{0Z}}}{\partial y^{2}}- \left(\frac{\partial \phi}{\partial y} \right)^2 A_{\mathrm{0Z}} & =
\ \mu_{0}  \varepsilon  \Bigg[  \frac{1}{\varepsilon} \frac{\partial \varepsilon}{\partial t}
\frac{\partial A_{\mathrm{0Z}}}{\partial t} + \frac{\partial^{2}A_{\mathrm{0Z}}}{\partial t^{2}}  \Bigg. \\  & \qquad \quad \left.  - \left(\frac{\partial \phi}
{\partial t}\right)^{2} A_{\mathrm{0Z}} \right],
\end{aligned}
\end{equation*}


and the other for the imaginary part:
\begin{equation}
\begin{aligned}
\frac{\partial^{2} \phi}{\partial y^{2}} A_{\mathrm{0Z}} + 2 \frac{\partial \phi}{\partial y} \frac{\partial
A_{\mathrm{0Z}}}{\partial y}  & = \ \mu_{0}  \varepsilon \left( \frac{1}{\varepsilon} \frac{\partial
\varepsilon}{\partial t} \frac{\partial \phi}{\partial t} A_{\mathrm{0Z}}  \right. \\ & \qquad  \left. + \frac{\partial^{2} \phi}{\partial
t^{2}} A_{\mathrm{0Z}} +2 \frac{\partial \phi}{\partial t} \frac{\partial A_{\mathrm{0Z}}}{\partial t}  \right).
\end{aligned}
\label{eq:amplitud}
\end{equation}


\subsection{The WKB approximation}
\qquad In order to solve for these variables, we first analyze the real part of the equation. If we assume that
$A_{\mathrm{0Z}}(y,t)$ varies slowly with position and time compared with $\phi(y,t)$ we can set
\begin{equation*}
\left( \frac{\partial \phi}{\partial y}\right)^2 A_{\mathrm{0Z}}  \gg   \frac{\partial^2 A_{\mathrm{0Z}}}{\partial y^2} ,
\end{equation*}

\begin{equation*}
\left( \frac{\partial \phi}{\partial t}\right)^2 A_{\mathrm{0Z}} \gg  \frac{1}{\varepsilon} \frac{\partial
\varepsilon}{\partial t} \frac{\partial A_{\mathrm{0Z}}}{\partial t}  +  \frac{\partial^2 A_{\mathrm{0Z}}}{\partial t^2}.
\end{equation*}

In this way we obtain the eikonal equation for $\phi(y,t)$,
\begin{equation}
\left( \frac{\partial \phi}{\partial y}\right)^2 - \frac{1}{v(t)^2} \left( \frac{\partial \phi}{\partial
t}\right)^2 = 0 ,
\label{eq:eikonal}
\end{equation}
where the following relation was used,
\begin{equation*}
\mu_{0} \ \varepsilon(t) = \frac{1}{v(t)^2}.
\end{equation*}

\subsection{The general solution and phase propagation}
\qquad Eq. (\ref{eq:eikonal}) can be solved by the characteristic method. We thus, rewrite this equation as

\begin{equation*}
\left ( \frac{\partial \phi}{\partial t} \right ) \pm  v(t)  \left ( \frac{\partial \phi}{\partial y} \right ) =0.
\end{equation*}

The general solution can be written as
\begin{equation*}
\phi(y,t) = \phi\left( W \right),
\end{equation*}
with $\phi$ an arbitrary function of the argument
\begin{equation}
W = W(y,t) = \pm  \ y - \int_{0}^{t} v(\tau) \ \mathrm{d} \tau.
\label{eq:Z}
\end{equation}

Replacing this solution in the amplitude eq. (\ref{eq:amplitud}) yields,

\begin{equation*}
\frac{\partial A_{\mathrm{0Z}}}{\partial t}  \pm   v(t)  \frac{\partial A_{\mathrm{0Z}}}{\partial y}  = \frac{1}{2} A_{\mathrm{0Z}} \frac{\partial }{\partial t} \left[ \ln v(t) \right].
\end{equation*}

The solution can be obtained via the characteristic method giving

\begin{equation*}
A_{\mathrm{0Z}}(y,t) = \sqrt{\frac{v(t)}{v_0}} \ A(W),
\end{equation*}
where $v_0$ is a constant with the dimension of velocity and $A(W)$ is an arbitrary function. For ease of notation, we have not distinguished between waves that propagate to increasing or decreasing values of $y$. This must be taken into account when solving for the propagation of the laser beam as light travels through different media.

Note also that both $\phi$ and $A$ depend on $W$. Thus one can write the solution as the real part of
\begin{equation*}
\mathrm{A_{z}} =  \sqrt{\frac{v(t)}{v_0}} A(W) \ \mathrm{e}^{\mathrm{i}\phi(W)},
\end{equation*}
with $A(W)$ and $\phi(W)$ real functions and where the phase represents a traveling wave with a time-dependent velocity. It is also useful to rewrite the solution as
\begin{equation}
\mathrm{A_{z}} =  \sqrt{\frac{v(t)}{v_0}} F(W) \ \mathrm{e}^{\mathrm{ i k} W},
\label{eq:Solucion_Full_A}
\end{equation}
with $F$ a real function and $\mathrm{k}$ a constant with dimension of inverse of length (and can be taken as the wave-number). This last form of the solution is used to obtain the electric and magnetic fields in the different media.


\section{Model of EOM, the transmitted fields and fundamental RAM}
\label{sec:Model_EOM}

\qquad The solution presented above can be used to obtain the propagation of light through media with different refractive indices. In particular, we are interested in describing an incoming laser in a media with constant index of refraction $n_0$ which then enters a region with a finite length and with a time-dependent index $n(t)$. After going through that region the laser then goes back to a region with constant index $n_0$. This step function model for the index of refraction describes the action of the EOM in a region of the crystal when the border effects of the EOM are not taken into account. To obtain the electromagnetic field in the three regions one imposes matching conditions at the incoming boundary $y =L_{1}$, at the outgoing boundary $y =L_{2}$, and then solve for the amplitudes of the fields.

We first write the electric field and the magnetic field in the three different regions as,
\begin{equation}
\mathbf{E}_{j}(y,t) = -
\frac{\partial \mathrm{A}_{j}}{\partial{t}} \  \mathbf{\hat{e}_{z}} ,
\label{eq:Ez}
\end{equation}

\begin{equation}
\mathbf{B}_{j}(y,t)=  \frac{\partial \mathrm{A}_{j}}{\partial{y}} \  \mathbf{\hat{e}_{x}}   ,
\label{eq:Bx}
\end{equation}
where the index $j$ take the following values $j=1,2,3$ depending on the region of interest. In our case the index $1$ identifies the incoming wave, $2$ the region with a time-dependent index of refraction and $3$ the outgoing region. Consequently we write the corresponding vector potential  using (\ref{eq:Solucion_Full_A}) as,

\begin{equation*}
\mathrm{A}_{1}(y,t)= F_{1}(W_{+}) \ \rm{e^{i \mathrm{k} W_{+}}} + F_{1}(W_{-}) \ \rm{e^{i \mathrm{k} W_{-}}},
\end{equation*}

\begin{equation*}
\mathrm{A}_{2}(y,t)= \sqrt{\frac{v(t)}{v_{0}}} \left[ F_{2}(W_{+}) \ \rm{e^{i \mathrm{k} W_{+}}} + F_{2}(W_{-}) \ \rm{e^{i \mathrm{k} W_{-}}} \right],
\end{equation*}

\begin{equation*}
\mathrm{A}_{3}(y,t)= F_{3}(W_{+}) \ \rm{e^{i \mathrm{k} W_{+}}},
\end{equation*}
and impose matching conditions for the electric and magnetic field at $y =L_{1}$ and $y =L_{2}$ to obtain the solution.

At this point we restrict ourselves to the discussion of laser light propagating through a crystal with an EOM whose dimension in the direction of laser propagation is smaller than that of the crystal  (see figure \ref{fig:figure1}). For this case, the index of refraction of the crystal in the EOM region is given by:

\begin{equation} 
n(t)= n_{0} \left[1 - \frac{\gamma}{2} \cos(\Omega t)\right] ,
\label{eq.n}
\end{equation}
where $n_{0}$ is the index of refraction of the crystal without the EOM and $\gamma$ is a very small
dimensionless parameter defined as,
\begin{equation*}
\gamma = n_{0}^2  \ r_{33} \frac{V_{0}}{d} ,
\end{equation*}
where $r_{33}$ is the electro-optic coefficient of the crystal, $V_{0}$ is the amplitude of
modulating voltage which generates electric field in the $\mathrm{"Z"}$ direction and $d$ is the
distance between the electrodes.

Thus, our EOM model to be solved is an incoming wave in a media with index $n_{0}$ entering a region with index $n(t)$ given by eq.(\ref{eq.n}) and then transmitted to a region with index $n_{0}$ again. 

Using the results presented in the Appendix \ref{Appendix A} we obtain the transmitted vector potential. For clarity, we rewrite this expression in the following form

\begin{equation}
\mathrm{A}_{3}(y,t) =  {A}_{0} \ \mathrm{e}^{\mathrm{i} \Phi \left( y,t \right)},
\label{eq:A3++}
\end{equation}
where
\begin{equation}\label{eq:phi3}
\Phi\left(y,t \right) = \mathrm{k} y - \omega_{0} t  - m \cos \left \{ \Omega \left[ \frac{y}{v_{0}} - \left( t-t_{0} \right) \right] \right \} ,
\end{equation}
and $m$ is the depth of the phase-modulation (modulation depth), defined as

\begin{equation}\label{eq:m_generalizado}
m = \frac{\omega_{0} \gamma}{\Omega} \ \sin\left( \frac{\Omega L}{2 v_{0}} \right).
\end{equation}

In (\ref{eq:phi3})  $t_{0}$ is a real constant given by

\begin{equation*}
t_{0} = \frac{L- 2 L_{2}}{2	v_{0}}.
\end{equation*}

Here $L=L_{2}-L_{1}$  represents the length of the electrode and $v_{0} = \tfrac{c}{n_{0}}$, where $c$ is the speed of light in vacuum.

This formulation shows that the vector potential emerging from the EOM is simply a phase-modulated wave with modulation depth $m$.

Since $(\frac{\Omega L}{2 v_{0}}) \ll 1$, we can approximate the modulation depth (\ref{eq:m_generalizado}) as

\begin{equation}
m =\frac{\gamma \omega_{0} L}{2 v_{0}} = \frac{\pi L}{\lambda_{0}} r_{33} \ n_{0}^{3} \frac{V_{0}}{d}.
\label{eq:def_modulation_depth}
\end{equation}

This last expression agrees with the usual definition of modulation depth that appears in the standard textbooks \cite{Haus}.

We rename the electric and magnetic fields emerging from the modulator region as $\mathbf{E}_{\mathrm{out}}(t) = \mathbf{E_{3}}(L_{2},t)$ and $\mathbf{B}_{\mathrm{out}} = \mathbf{B_{3}}(L_{2},t)$, respectively.

 Replacing  the vector potential (\ref{eq:A3++}) in (\ref{eq:Ez}) we obtain


\begin{equation}\label{eq:Eout}
\mathbf{E}_{\mathrm{out}}(t) = \  \mathrm{i}  {E}_{0}  \left[ 1 - \frac{m \Omega}{\omega_{0}} 
\sin\left( \Omega  t  \right) \right]  \mathrm{e}^{ \mathrm{i} \left[ \mathrm{k} L_{2} - \omega_{0} t - m \cos(\Omega t) \right] } \ \mathbf{\hat{e}_{z}},  
\end{equation}

with ${E}_{0}= {A}_{0}\ \omega_{0}$ and $m$ is given by (\ref{eq:def_modulation_depth}).

Similarly, using (\ref{eq:Bx}) the transmitted magnetic field results

\begin{equation}\label{eq:Bout}
\mathbf{B}_{\mathrm{out}}(t) =  \ \frac{\mathrm{i} {E}_{0}}{v_{0}}  \left[ 1 - \frac{m \Omega}{\omega_{0}}\sin \left( \Omega t \right)\right]  \mathrm{e}^{ \mathrm{i} \left[ \mathrm{k} L_{2} - \omega_{0} t - m \cos(\Omega t) \right] } \ \mathbf{\hat{e}_{x}}. 
\end{equation}

As one can see, the electric and magnetic fields emerging from the EOM have a $\mathrm{RAM}$ at frequency $\Omega$. This is rather surprising since it is usually assumed that only the phase is affected by the EOM phase modulation. Indeed, expression (\ref{eq:Eout}) contradicts equations on EOM phase modulation from standard textbooks. (Please see page 333, eq. (12.19) of ref. \cite{Haus} or also page 244, eq. (7.3-14) of ref. \cite{YarivOWC}). The textbook equation for the electric field emerging from an electro-optic modulator takes a form free from $\mathrm{RAM}$. A direct consequence of this result is a modulation of the outgoing intensity of a laser beam after it goes through a region with an EOM.

{
Since the emerging beam of the crystal illuminates a photodiode, the resulting photocurrent ($i_{c}$) is given by
\begin{equation}
i_{c}(t) \propto  \mathrm{E}_{\mathrm{out}}(t) \ \mathrm{E}_{\mathrm{out}}^{\ast}(t),
\label{eq:def_photocurrent}
\end{equation} 
}

Replacing (\ref{eq:Eout}) in (\ref{eq:def_photocurrent}) we obtain 

\begin{equation}
i_{c}(t)  \propto  {E_{0}^{2}} \left[ 1 - \frac{2 m \Omega}{\omega_{0}}
\sin \left( \Omega t \right) \right],
\label{eq:current}
\end{equation}
where it is clear that there exists $\mathrm{RAM}$  in the intensity emerging of EOM and  it is used to define $\mathrm{RAM_{F}}$.

The $\mathrm{RAM_{F}}$ is a periodic modulation of the transmitted wave intensity with frequency $\Omega$ for an ideal EOM with other sources of $\mathrm{RAM}$ being omitted. 
In the Appendix \ref{Appendix C} we compute $\mathrm{RAM_{F}}$:

\begin{equation}
\mathrm{RAM_{F}} =  \frac{2 m \Omega}{\omega_{0}}.
\label{eq:FLN}
\end{equation}

Table \ref{tab:tabla1} gives the $\mathrm{RAM_{F}}$ levels for the EOM of  aLIGO. Note that  $\omega_{0} =  \frac{ 2 \pi c}{\lambda_{0}} $, $ \lambda_{0}=1064\,$nm and $\Omega= 2 \pi f_{m}$, where $f_{m}$  is the modulation frequency of the EOM.

\begin{table}[htbp]
\centering
\caption{\bf $\mathrm{RAM_{F}}$ levels calculated from (\ref{eq:FLN}) for the EOM of aLIGO}
\begin{tabular}{ccccc}
\hline
${m}$  &${f_{m}}$ [MHz] & $\mathrm{RAM}$  & $\mathrm{RAM_{F}}$  & $\eta = \frac{2 \Omega}{\omega_{0}}$\\
\hline
$0.15$ &$45.3$ & $6.2\times 10^{-6}$ & $4.8\times 10^{-8}$ &$3.2 \times 10^{-7}$ \\			
$0.39$ &$9.18$ & $3.9\times 10^{-5}$ & $2.5\times 10^{-8}$ &$6.5 \times 10^{-8}$  \\
$0.09$ &$24.0$ & $1.0\times 10^{-6}$ & $1.5\times 10^{-8}$ &$1.7 \times 10^{-7}$  \\ 
\hline
\end{tabular}
\label{tab:tabla1}
\end{table}

In table \ref{tab:tabla1},  the third column shows the experimental values of $\mathrm{RAM}$ for aLIGO at Livingston LIGO Observatory (LLO) \footnote{\url{https://dcc.ligo.org/public/0108/E1300758/001/E1300758-v1.pdf}} and the fifth column shows $\eta= \frac{\mathrm{RAM_{F}}}{m}$, the ratio between the $\mathrm{RAM_{F}}$ and the modulation depth $m$. Using  (\ref{eq:FLN}) gives
$\eta = \frac{2 \Omega}{\omega_{0}} $.



\section{Fundamental RAM and Conservation of energy}
\label{sec:micro_macro}

\qquad In this section we show that $\mathrm{RAM_{F}}$ can also be obtained from conservation of energy if one introduces a more general version of the Poynting theorem when the electrical permittivity of the media is time-dependent. We also show that the same physical principles that yield the 
propagation of light inside the crystal of EOM can be used to describe $\mathrm{RAM_{F}}$.

It follows from the Faraday and Ampere-Maxwell laws (\ref{eq:Faraday}) and (\ref{eq:Ampere-Maxwell}) that we can write down the electromagnetic energy flux associated with a volume $V$ and its boundary $\partial V$ as:

\begin{equation}{\label{eq:Poynting_generalizado}}
\oint_{\partial V} \mathbf{S \cdot \hat{n}} \ \mathrm{d}a =-\int_{V} \left(  \mathbf{E \ \cdot} \ \frac{\partial \mathbf{D}}{\partial t} + \mathbf{H \ \cdot} \ \frac{\partial \mathbf{B} }{\partial t}  \right)  \mathrm{d}V,
\end{equation}
where $\mathbf{S} = \frac{1}{\mu_0} \left( \mathbf{E} \times \mathbf{B}\right)$ is the Poynting vector,  and with $V$ the EOM volume.   Using the relations (\ref{eq:BH}) and (\ref{eq:ED}) 
and taking the time average of  (\ref{eq:Poynting_generalizado}) in the adiabatic approximation on a given period of the incoming electromagnetic wave, $T= \frac{2\pi}{\omega_{0}}$, a straightforward calculation leads to

\begin{equation}
\oint_{\partial V} \left \langle \mathbf{S \cdot \hat{n}} \right \rangle \ \mathrm{d}a =-   \int_{V}   \left( \left \langle \frac{\partial \mathrm{u_{em}}}{\partial t}  \right \rangle  + \left \langle  \frac{1}{2} \frac{\partial \varepsilon}{\partial t} \ \mathrm{E}^{2} \right \rangle \right) \mathrm{d}V ,
\label{eq:poyntingeneral}
\end{equation}
where
\begin{equation*}
 \mathrm{u}_{\mathrm{em}} \equiv \frac{\varepsilon(t)}{2} \  \mathrm{E}^{2} + \frac{1}{2 \mu_{0}} \ \mathrm{B}^{2},
\end{equation*}
is the (generalized) energy density of the electromagnetic field and the symbol $ \left \langle  \bullet \right \rangle  $ labels the time average over one cycle of the carrier.

The first term on the r.h.s. of  (\ref{eq:poyntingeneral}) represents the average power of the electromagnetic field, and the second term on the r.h.s. of (\ref{eq:poyntingeneral}) depends on the variation rate of the electrical permittivity.

At this point we can address several issues:

\begin{itemize}
\item Note that when $\varepsilon$ does not depend on the time we recover the usual expression for the conservation of energy.
\item It would appear that this first term on the r.h.s. gives a bigger contribution to the Poynting flux than the second term. However, each term gives an identical contribution to the equation. This follows from the fact that the magnetic contribution and the static part of the electric contribution to the energy density yield a vanishing flux. Hence, only the time-dependent part of the energy density matters and gives a similar contribution to the second term of the equation.
\item If we compute the flux of the Poynting vector for our model we obtain

\begin{equation}
\oint_{\partial V} \left \langle \mathbf{S \cdot \hat{n}} \right \rangle \ \mathrm{d}a = - \frac{E_{0}^{2}}{2 \mu_{0} v_{0}} l_{x} l_{z}  \ \mathrm{RAM_{F}} \ \sin\left( \Omega t  \right),
\label{eq:Flujo_neto}
\end{equation}

where $l_{x}$ and $l_{z}$ are the crystal dimensions perpendicular to the electromagnetic wave.

Therefore, from the macroscopic point of view, $\mathrm{RAM_{F}}$ is related to an external flux of energy 
ingoing to the crystal. Since modulating field constant ($\Omega=0$) implies null $\mathrm{RAM_{F}}$, then $
\mathrm{RAM_{F}}$ is a dynamic effect which is a consequence of the work done by the variable modulating field on the 
media.
\end{itemize}

\section{Analysis of cancellation of fundamental RAM}
\label{sec:cancellation_RAM}
\qquad Since the $\mathrm{RAM_{F}}$ effect is fully predictable, it is natural to think about its total cancellation as an affordable goal.

It is worth noting that the intensity of the emerging field of the EOM can be affected by two types of $\mathrm{RAM}$: $\mathrm{RAM_{S}}$ (systematic $\mathrm{RAM}$) and $\mathrm{RAM_{N}}$ (noise $\mathrm{RAM}$). 

$\mathrm{RAM_{S}}$ represents all those $\mathrm{RAM}$ sources that are predictable. This type of sources includes $\mathrm{RAM_{F}}$ and, for example, the constant misalignment angles of the EOM optical configuration. 

$\mathrm{RAM_{F}}$ is an absolutely foreseeable component of $\mathrm{RAM}$ which we have named it as fundamental because it is an unavoidable consequence of the phase modulation process itself. In other words, this component of $\mathrm{RAM}$ is called fundamental because even in the ideal case where all the other possible sources of $\mathrm{RAM}$ are eliminated, $\mathrm{RAM_{F}}$ would always be present. 

$\mathrm{RAM_{N}}$ instead is an unpredictable parasitic component of $\mathrm{RAM}$ arising from several random sources such as temperature fluctuations, mechanical vibrations, etalon effects due to defects of the crystal, photorefractive effect, etc. In the usual experimental situations, $\mathrm{RAM_{N}}$ is much greater 
than $\mathrm{RAM_{F}}$.

Since until now the existence of $\mathrm{RAM_{F}}$ was unknown, efforts have been addressed to remove only $\mathrm{RAM_{N}}$. Thus, depending on the specific field of experimental application, different active feedback control schemes were proposed to remove or reduce $\mathrm{RAM_{N}}$ (\cite{Wong1985}, \cite{Zhang2014}, \cite{Li2014}).
In \cite{Wong1985}, for spectroscopy applications, the authors claim to reduce $\mathrm{RAM_{N}}$ up to shot noise level.

In optical-cavity-based frequency ultra-stabilization, different closed-loop control schemes were proposed (\cite{Zhang2014} , \cite{Li2014}). In \cite{Zhang2014}, 
the active control implemented in wave-guide EOM managed to reduce the $\mathrm{RAM_{N}}$  to limit it to maximum values of $5 \times 10^{-6}$ and the Allan deviation associated with $\mathrm{RAM_{N}}$ fluctuation was limited to $10^{-6}$ for average time between $\mathrm{1-1000 \ s} $. In \cite{Li2014}, the authors claim to have reached a minimum value of the Allan deviation of $\mathrm{RAM}$ of $2 \times 10^{-7}$ for an average time of $2$s. It should be noted that $\mathrm{RAM_{F}}$ does not affect the value of Allan deviation of $\mathrm{RAM}$ since it represents the constant term of the sideband amplitudes.  
 
It is important to keep in mind that the different attempts to eliminate $\mathrm{RAM}$ are based on a model where the intensity of the emerging EOM laser is only affected by a random parasitic misalignment between the polarization of the laser and the main axes of the crystal. This misalignment causes two beams with 
orthogonal polarization, ordinary $(o)$ and extraordinary $(e)$. Due to the birefringence of the crystal, both beams travel with different speeds, and the superposition of both beams in the polarizer placed at the output of the EOM then generates $\mathrm{RAM}$. This model does not include $\mathrm{RAM_{F}}$ since its existence was unknown. Therefore, we can conjecture that $\mathrm{RAM}$, i.e. $\mathrm{RAM_{N}}$ and $\mathrm{RAM_{S}}$, should be able to be removed, provided that the feedback control system is based on an electromagnetic propagation model that includes both  $\mathrm{RAM}$ sources. Thus, having the theoretical formula from de emerging beam of intensity that includes $\mathrm{RAM_{N}}$ and $\mathrm{RAM_{S}}$ will be helpful to understand the minimum experimental values of $\mathrm{RAM_{N}}$ obtained in \cite{Zhang2014} and \cite{Li2014}, and to determine the optimal sensing and control schemes that will allow cancellation of the $\mathrm{RAM}$.

\begin{figure}[htbp]
\centering
\includegraphics[width=\linewidth]{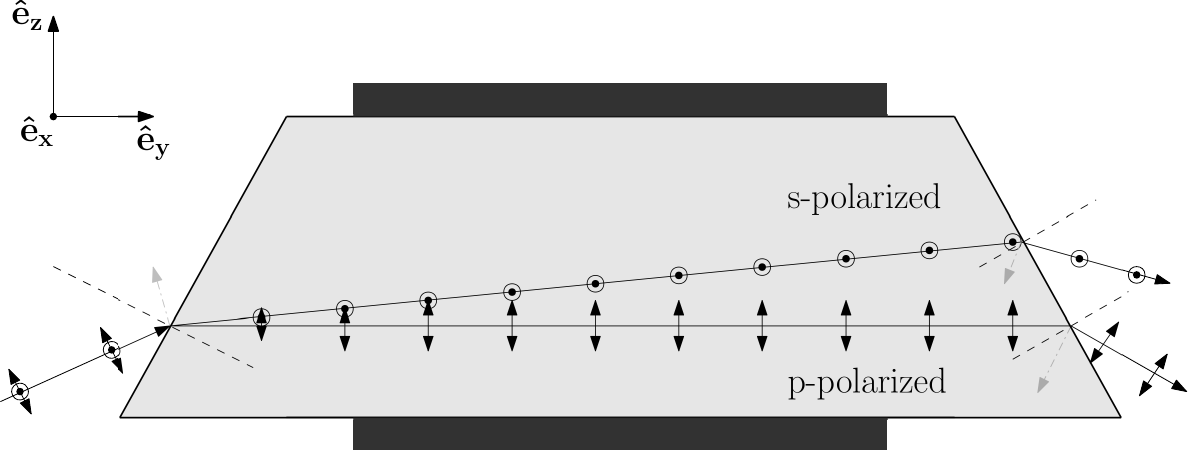}
\caption{Separation of the polarizations in a wedge crystal.}	
\label{fig:split_polarization}
\end{figure}

On the other hand, the situation is different for the wedge-shaped crystal EOM analyzed in the present work. 
In fact, the wedge shape of the crystal separates the directions of the rays $o$ and $e$ when they emerge from the crystal, see fig. (\ref{fig:split_polarization}). 

This avoids contributions to $\mathrm{RAM}$ arising from both the main etalon effect and the misalignment. For this reason, expression (\ref{eq:FLN}) of this work is valid for the wedge-shaped crystal EOM, 
which does not correspond to the experimental situations analyzed in \cite{Wong1985}, \cite{Zhang2014} and 
\cite{Li2014}.

\section{Conclusions}
\label{sec:conclusions}

\qquad In this work, we have established a lower limit for the $\mathrm{RAM_{F}}$ effect when using electro-optic modulators. We showed that when some region inside a crystal has a time-dependent index of refraction produced by EOM, the phase and amplitudes of the outgoing electromagnetic fields are modulated with the frequency of the EOM.
{This limit was obtained using an approximation to Maxwell's equations and a particular model for the propagation media.
We assume that the crystal permittivity is time-dependent inside the modulation zone (between the pair of electrodes) and is constant outside of it, so it is spatially piecewise homogeneous $\varepsilon(t)$}. Furthermore, we assume that $\varepsilon(t)$ is a slowly varying function of time with its angular frequency $\Omega$ much smaller than the corresponding frequency $\omega_{0}$ of the laser wave propagating in the crystal ($\frac{\Omega}{\omega_{0}} \approx 1 \times 10^{-7}$). In this situation, we can apply the so-called WKB approximation and derive effective equations of motion for the propagating wave. The plane wavefront assumption simplifies the analysis but does not change the general result since the WKB equations are linear in the electric and magnetic fields and we are simply taking a Fourier decomposition.

We have not considered border effects on the EOM either, i.e., the modulating electric field is assumed to be homogeneous even at both ends of the modulation electrodes and null outside of them. So the interfaces in which the laser enters and leaves the modulation zone to become discontinuity planes of the refractive index. Furthermore, we have restricted ourselves to consider wedge-shaped crystal EOM whose oblique faces lie outside of the modulation zone (see fig. \ref{fig:figure2}). This does not represent a restrictive condition since it is usual in experimental layouts of EOM. For those reasons, the oblique interfaces separate two constant refractive index media (air/crystal without modulation field or vice versa). Consequently, when the laser goes through these interfaces no contribution to $\mathrm{RAM_{F}}$ can be produced neither on the reflected nor on the transmitted components of the laser beam. This means that $\mathrm{RAM_{F}}$ is strictly generated by modulation zone.

We could have taken border effect into account by substituting the step function model for $n$ with a smooth function that takes into account the border effects of the EOM. We claim that $\mathrm{RAM_{F}}$ effect is also present when a more general configuration is adopted.

On the other hand, it would appear that we have neglected the etalon effect without justification. However, this is not so. As it is well known, this effect is generated by the contribution of the successive internal reflections to the emerging laser. Two types of interfaces generate internal reflections: the oblique ones 
(air/crystal without modulating field or vice versa) and the vertical ones (crystal without modulating field/crystal with the modulating field, or vice versa). The reflections at the oblique faces do not contribute to the outgoing beams because they are deliberately deflected away from the path of the main beam. The reflections at the two vertical interfaces are crucial for the calculation of the emerging fields of the crystal. However, it is important to note the jump of the refractive index through both surfaces is of the first order in $\gamma$, with $\gamma \ll 1$ (see(\ref{eq.n})). So from the second reflection onwards, the amplitudes of the reflective fields are $\gamma^{l}$-order with $l \geqslant 2$, and they are out of our approximation order. 

Based on the above considerations, we claim it is impossible to obtain a pure phase modulation without an associated $\mathrm{RAM_{F}}$ effect.
The order of magnitude of $\mathrm{RAM_{F}}$ obtained ranges from $10^{-7}$ to $10^{-8}$, depending on the modulation frequency $\Omega$, carrier laser frequency $\omega_{0}$ and the phase modulation depth $m$, as it is shown in table \ref{tab:tabla1}.
Since one of the main goals of this work is to understand the $\mathrm{RAM}$ effect on the laser beam of aLIGO, it is instructive to compare our results to the experimental values of $\mathrm{RAM}$ in aLIGO, which range from $10^{-5}$ to $10^{-6}$.
Although this work shows that it is impossible to eliminate the $\mathrm{RAM}$ effect in the beam intensity, a meticulous search of other sources may help to reduce the present level of $\mathrm{RAM}$.
It is worth mentioning that the $\mathrm{RAM}$ level predicted in this work is also present in more general configurations. As we said before, a wedge-shaped crystal avoids the etalon effect so the transmitted laser beam in the crystal is not subject to multiple reflections on the wedged boundary. As a consequence of this shape of the crystal, it is clear to see that in all experimental EOM layouts, every single EOM can be modeled as we did in section \ref{sec:Model_EOM}. For these reasons $\mathrm{RAM_{F}}$ obtained in this work will be presented in all experimental configurations of EOM. As an example, the aLIGO EOM layout is a wedge-shaped RTP crystal over which three consecutively coupled of an electrode are placed. Each pair of electrodes is fed with AC voltage needed to generate three modulating fields of different frequencies. Therefore, the emerging laser beam of this series of modulators undergoes three-phase modulation processes, in three different frequencies.  The value of each frequency is chosen to control the length of three different cavities. Our result for this case implies that, regardless of the particular position inside the series, each modulator generates a $\mathrm{RAM_{F}}$ in laser, at different frequency, which can be calculated using (\ref{eq:FLN}).
It is also important to note that our result implies $\mathrm{RAM_{F}}$ will be present in all experimental applications that use EOMs, for example, frequency-modulation spectroscopy and optical-cavity-based frequency ultra-stabilization. 
Moreover, we conjecture that $\mathrm{RAM_{F}}$ will exist in any phase modulation process where an external agent (the modulating electric field for the crystal-based EOM) produces changes in the refractive index of the medium where the laser travels. As an example, we can mention the acoustic optical modulators and the waveguide
based EOM. Indeed, $\mathrm{RAM_{F}}$ will be a consequence of the work done by the external agent on the medium in which the laser propagates.

We want to emphasize that, although $\mathrm{RAM_{F}}$ will always be present in phase modulation processes with 
EOMs,  in general, the value of $\mathrm{RAM_{F}}$  depends on the specific optical layout of the EOM.

Finally, we have analyzed the suppression of $\mathrm{RAM_{F}}$. In spite of the small magnitude 
of $\mathrm{RAM_{F}}$, we conjecture that all types of $\mathrm{RAM}$, 
including $\mathrm{RAM_{F}}$, should be able to be removed, provided that the feedback control system is based 
on an electromagnetic propagation model that includes both $\mathrm{RAM}$ sources.


\appendix

\section{The solutions to the WKB equations in a crystal of the EOM} \label{Appendix A}
\qquad Following the steps outlined in section \ref{sec:Model_EOM}, we write down the solutions to the WKB equations in the three regions of interest. There are in principle four unknown functions of W that should be fixed by the matching conditions, i.e., they provide 4 equations for 4 unknowns functions of time. Since the idea is to solve them for a quasi-static configuration, it is worth noting that it should be similar to the static situation, $v = const.$ We thus obtain first the solution for this familiar case and then generalize to the quasi-static model.
\subsection{The static situation}

\qquad The three media have refractive indices given by $n_{1}=n_{0}$,  $n_{2}=n_{0}(1- \frac{\gamma}{2})$ , $n_{3}=n_{0}$, with interfaces at $y=L_{1}$ and $y=L_{2}$. In this case we have the known result:
\begin{equation}\label{eq:CE_A1}
\mathbf{A_{1}}(y,t)= \left[  A_{1+} \ \mathrm{e}^{\mathrm{i} ( \mathrm{k}_{1} y - \omega_{0}t )} + A_{1-} \ \mathrm{e}^{-\mathrm{i} (\mathrm{k}_{1} y + \omega_{0}t )} \right] \ \mathbf{\hat{e}_{z}}  ,
\end{equation}
\begin{equation}\label{eq:CE_A2}
\mathbf{A_{2}}(y,t)= \left[  A_{2+} \ \mathrm{e}^{\mathrm{i} ( \mathrm{k}_{2} y - \omega_{0}t )} + A_{2-} \ \mathrm{e}^{-\mathrm{i} ( \mathrm{k}_{2} y + \omega_{0}t )} \right] \ \mathbf{\hat{e}_{z}}  ,
\end{equation}
\begin{equation}\label{eq:CE_A3}
\mathbf{A_{3}}(y,t)=   A_{3+} \ \mathrm{e}^{\mathrm{i} ( \mathrm{k}_{3} y - \omega_{0}t )}  \ \mathbf{\hat{e}_{z}}  ,
\end{equation}
where $A_{j\pm}$ are constants that are fixed from the matching conditions. The wavenumber is given by $\mathrm{k}_{j} =\frac{ n_{j} \omega_{0}}{c}$,  and $j=1,2,3$ labels the three refractive indices.

After solving the matching conditions one obtains,
\small

\begin{equation}
\begin{aligned}
\mathbf{A_{1}}(y,t) = & \ A_{0} \ \mathrm{exp}{\left[ \mathrm{i} \left( \mathrm{k} y - \omega_{0} t \right) \right]} \ \mathbf{\hat{e}_{z}} \   \\ & + \frac{\gamma}{4} \ A_{0} \ 
\mathrm{exp}{\left[ - \mathrm{i}\left( \mathrm{k} y + \omega_{0} t - 2 \mathrm{k} L_{1} \right) \right]} \ \mathbf{\hat{e}_{z}} 
\\ & - \frac{\gamma}{4} \ A_{0} \  \mathrm{exp}{ \left \{ - \mathrm{i} \left[ \mathrm{k} y + \omega_{0} t  - 2 \mathrm{k} L_{2}  + \mathrm{k} \gamma L \right] \right \} }  \ \mathbf{\hat{e}_{z}},
\label{eq:E1}
\end{aligned}
\end{equation}

\begin{equation}
\begin{aligned}
\mathbf{A_{2}}(y,t) = & \  A_{0}  \left(1+\frac{\gamma}{4}\right)  \ \mathrm{exp}
{\left \{ \mathrm{i}\left[ \mathrm{k} y  - \omega_{0} t  - \frac{\mathrm{k} \gamma}{2} \left(y - L_{1} \right) \right] \right \} } \ \mathbf{\hat{e}_{z}} \\ & -\frac{\gamma}{4} \ A_{0} \ \mathrm{exp}{\left \{ -\mathrm{i}\left[ \mathrm{k} y + \omega_{0} t - \frac{ \mathrm{k} \gamma}{2} \left( y + L_{1} \right) \right. \right.} \\ & \qquad \qquad \qquad + {\left. \left. \mathrm{k} L_{2}(\gamma
	- 2)\right] \right \} } \  \mathbf{\hat{e}_{z}},
\end{aligned}
\end{equation}


\begin{equation}
\mathbf{A_{3}}(y,t) = A_{0} \ \mathrm{exp}{\left \{ \mathrm{i} \left[ \mathrm{k} y - \omega_{0} t - \frac{ \mathrm{k} \gamma L}{2} \right] \right \} } \
\mathbf{\hat{e}_{z}},
\label{eq:E3}
\end{equation}
\normalsize

where $\mathrm{k} = \frac{\omega_{0}}{v_{0}}$ and $L = L_{2} - L_{1}$.

\subsection{The non-static case}
\label{subsec:matching}
\qquad Solving the matching condition for the non-static case yields the following results:

\small
\begin{equation}
F_{1}(W_{+}) \ \mathrm{e}^{ \mathrm{i k} W_{+}}=A_{0}\ \mathrm{exp}{\left[ \mathrm{i} \left( \mathrm{k} y -\omega_{0}t\right) \right]},
\label{eq:A1+}
\end{equation}

\begin{equation}
\begin{aligned}
F_{1}(W_{-}) \ \mathrm{e}^{\mathrm{i k} W_{-}} = & \  \frac{\gamma}{4} \ A_{0} \  \cos\left[\Omega
\left( t - \frac{L_{1} - y}{v_{0}} \right)\right]  \\ & \times \mathrm{exp}{\left[- \mathrm{i} \left( \mathrm{k} y +\omega_{0} t - 2 \mathrm{k} L_{1} \right) \right]} \\ & - \frac{\gamma}{4} \
A_{0} \  \cos\left[ \Omega \left( t - \frac{ L_{2} - y}{v_{0}} \right)\right]  \\ & \times 
\mathrm{exp}{ \Bigg( -\mathrm{i} \left \{  \mathrm{k} y + \omega_{0} t - 2 \mathrm{k} L_{2}  \right. \Big.} \\ &    \qquad \quad { + \frac{\omega_{0} \gamma}{\Omega} \sin\left(\frac{\Omega L}{v_{0}} \right)} \\ & \qquad \quad  { \Bigg. \left. \times   \cos \left[\Omega \left( t - \frac{ L_{2} - y}{v_0} \right)\right]  \right \}  \Bigg)},
\end{aligned}
\label{eq:A1-}
\end{equation}


\begin{equation}
\begin{aligned}
\sqrt{\frac{v(t)}{v_{0}}} \ F_{2}(W_{+}) \ \mathrm{e}^{\mathrm{i k} W_{+}} & = \ A_{0} \left[ 1+ \frac{\gamma}{4} \cos(\Omega t)\right] \\ & \times \mathrm{exp}{\Bigg( \mathrm{i} \left \{ \mathrm{k} y - \omega_{0} t \right. \Bigg.} \\ & \qquad  
- {\frac{\omega_{0} \gamma}{\Omega} \sin\left[\frac{\Omega \left( y - L_{1} \right)}{2 v_{0}} \right] }   \\ & { \qquad \times \Bigg. \left. \cos\left[ \Omega \left( t - \frac{y - L_{1}}{2 v_{0}} \right)\right] \right \} \Bigg)},
\end{aligned}
\label{eq:A2+}
\end{equation}


\begin{equation}
\begin{aligned}
\sqrt{\frac{v(t)}{v_{0}}} \  F_{2}(W_{-}) \ \mathrm{e}^{\mathrm{i
		k} W_{-}} & = - \frac{\gamma}{4} \ A_{0}  \  \cos\left[ \Omega
\left( t + \frac{y - L_{2}}{v_{0}} \right)\right] \\  &
\times \mathrm{exp}{\Bigg( -\mathrm{i} \left \{ \mathrm{k} y  + \omega_{0} t  - 2 \mathrm{k} L_{2} \right. \Bigg.} \\ & \qquad + {\frac{\omega_{0} \gamma}{\Omega} 
	\sin \left[ \frac{ \Omega \left( 2 L_{2} - L_{1} - y \right)}{2 v_{0}}  \right]} \\ & {  \times \Bigg. \left.  \cos\left[\Omega
	\left( t - \frac{\Omega \left(2 L_{2} - L_{1} - y \right)}{2 v_0} \right)\right] \right \} \Bigg) },
\end{aligned}
\label{eq:A2-}
\end{equation}


\begin{equation}
\begin{aligned}
F_{3}(W_{+}) \ \mathrm{e}^{ \mathrm{i k} W_{+}}= & \ A_{0} \ \mathrm{exp}{\Bigg(  \mathrm{i} \left \{ {\mathrm{k} y - \omega_{0} t - \frac{\omega_{0} \gamma}{\Omega} \sin \left( \frac{\Omega L}{2 v_{0}} \right)} \right. \Bigg.} \\  & \qquad {\Bigg. \left. \times   \cos\left[ \Omega \left( t - \frac{L - 2 L_{2} + 2 y }{2 v_{0}} \right)\right] \right \} \Bigg) }.
\end{aligned}
\label{eq:A3+}
\end{equation}


\normalsize
As one can check, the solutions $\mathrm{A}_{j}(y,t)$, satisfy the matching conditions and converge to the static case when the limit $\Omega \to 0$ is taken.


\section{An alternative method: following the wave} \label{Appendix B}

\qquad In this Appendix we reobtain the solution presented in Appendix \ref{Appendix A} corresponding to the three media with indices of refraction: $n_{1}=n_{0}$, $n_{2}(t)= n_{0} \left[1- \frac{\gamma}{2} \cos(\Omega t) \right]$ and $n_{3}=n_{0}$ using a different ansatz.  This method is based on ray tracing and considers the successive  transmissions and reflections of the original incoming wave as it goes through the two interfaces between the three media.

As we have seen in the previous section, the  WKB approximation can be used to obtain an explicit form of the vector potential $\mathrm{A_{z}}(y,t)=A_{\mathrm{0Z}}(W) \ \mathrm{e}^{\mathrm{i}\phi(W)}$, with $W$ given in eq.(\ref{eq:Z}). Assuming the phase $\phi(W)$ to be an increasing function of $W$, we define

\begin{equation}
\mathrm{k}(y,t)\equiv \left(\frac{\partial \phi}{\partial y} \right)_{t},
\label{eq:k}
\end{equation}
\begin{equation}
\omega(y,t)\equiv - \left(\frac{\partial \phi}{\partial t}\right)_{y},
\label{eq:omega}
\end{equation}
as the generalized wavenumber and frequency respectively. Inserting the definitions (\ref{eq:k}) and (\ref{eq:omega}) in (\ref{eq:eikonal}) yields

\begin{equation*}
\mathrm{k}(y,t) = \pm \frac{\omega(y,t)}{v(t)}.
\end{equation*}
In the above equation $\pm$ means a wave propagating to increasing or decreasing values of $y$.

\begin{figure}[htbp]
\centering
\includegraphics[width=\linewidth]{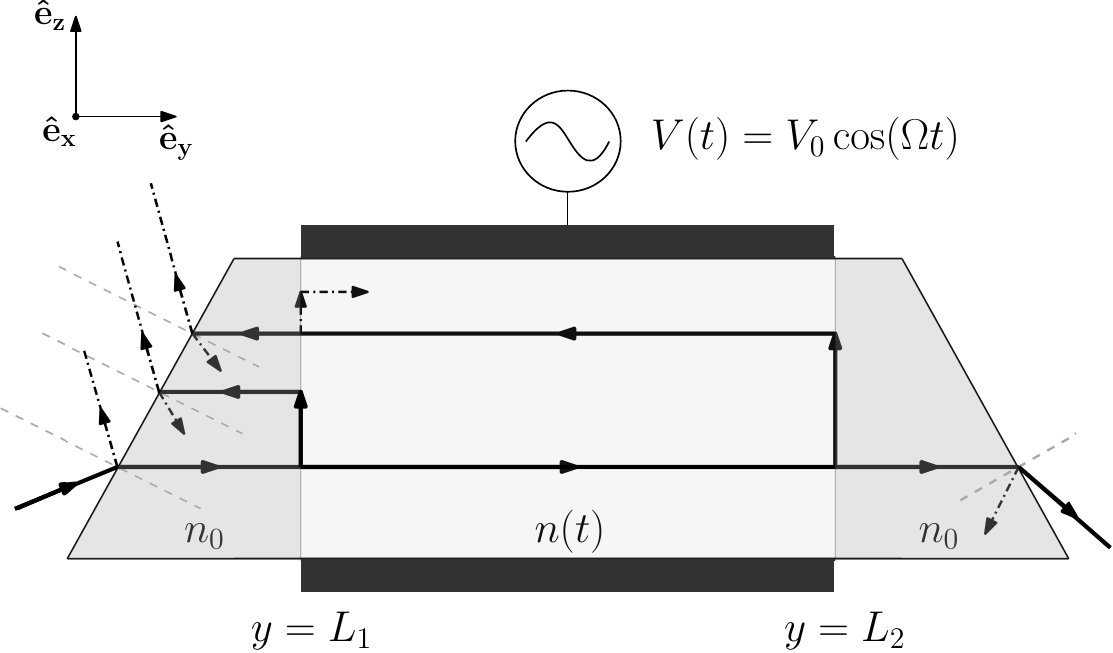}
\caption{Ray tracing in the EOM with wedge-shape crystal.}	
\label{fig:figure1}
\end{figure}

\begin{figure}[htbp]
\centering
\includegraphics[width=\linewidth]{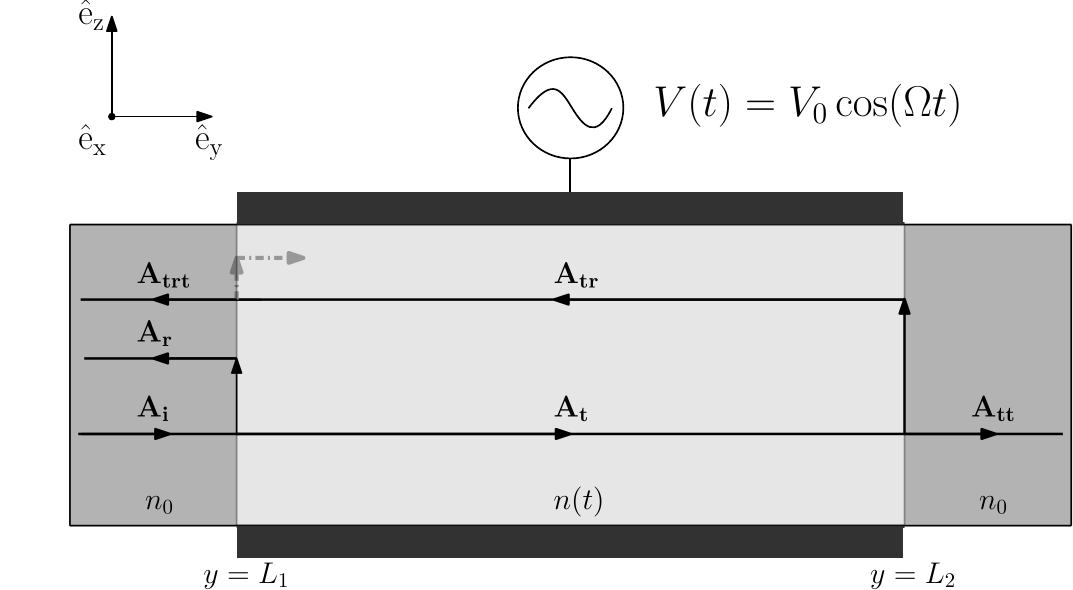}
\caption{Model for EOM.}
\label{fig:figure2}
\end{figure}

In principle, imposing the matching conditions for  $\mathbf{E}$ and $\mathbf{B}$ at $y=L_{1}$ and $y=L_{2}$ yields an infinite set of waves going back and forth which contribute to the final form of the fields in the three regions. However, as one follows the original incoming ray and assumes $\gamma \ll 1$, it is clear that the second and third reflection inside the media $n_{2}(t)$ will be proportional to the second and third power of $\gamma$ respectively.
We thus assume that only the rays that contribute to linear order in $\gamma$ are relevant for the calculation (see figure \ref{fig:figure1} and figure \ref{fig:figure2}).

These propagating fields are: $\mathbf{A_{i}}$ the incident vector potential in media $n_{1}$, $\mathbf{A_{r}}$ the reflected potential at $L_1$ propagating in media $n_{1}$, $\mathbf{A_{t}}$ the transmitted vector potential to the media $n_{2}$,  $\mathbf{A_{tr}}$ the reflected potential at $L_2$ propagating in media $n_{2}$, $\mathbf{A_{trt}}$ the transmitted wave to the media $n_{1}$ after it was reflected at $L_2$. Finally,  $\mathbf{A_{tt}}$ is the transmitted potential propagating in $n_{3}$.  Note that the reflected wave in $n_{1}$ is formed by the addition of two rays that are linear in  $\gamma$: $\mathbf{A_{r}}$ and $\mathbf{A_{trt}}$.

In the static case, we assume the angular frequency is the same for all the waves propagating in the three media. However, in the quasi-static case, the angular frequencies differ from $\omega_{0}$. After the calculation is done we must check that in the static limit, ($\Omega \to 0$), all the angular frequencies converge to $\omega_{0}$.

As an example, we consider the different fields that are traveling back in media $n_{1}$.
$\mathbf{A_{r}}$ has a constant angular frequency since it does not enter the time-dependent region, whereas $\mathbf{A_{trt}}$ has an angular frequency $\omega_{trt}(y,t)$ since it contains information of the transit time in the modulated media.

Motivated by the above example we introduce the following ansatz: \textit{"the transmitted and reflected waves at a given interface, that are generated by the same incident wave, possess the same angular frequency as the latter"}. The condition imposed on the angular frequencies at a given interface yields a matching condition for the phase of the incoming and outgoing fields. It follows from (\ref{eq:omega}), that these conditions fix the phase $\phi$ up to a constant.

When $\mathbf{E_{i}}$ reaches $y=L_{1}$ a transmitted $\mathbf{E_{t}}$  and reflected $\mathbf{E_{r}}$ fields are produced. The matching conditions for the angular frequencies are:
\begin{equation*}
\omega_{i}(L_{1}, t)= \omega_{t}(L_{1}, t) =\omega_{r}(L_{1}, t)  =\omega_{0},
\end{equation*}
together with the conditions for the amplitude of the fields:
\begin{equation*}
\mathbf{E_{i}} \left( L_{1},t \right) + \mathbf{E_{r}} \left( L_{1},t \right)  -
\mathbf{E_{t}}\left( L_{1},t \right)  =0,
\end{equation*}
\begin{equation*}
\mathbf{B_{i}} \left( L_{1},t \right) + \mathbf{B_{r}} \left( L_{1},t \right)  -
\mathbf{B_{t}}\left( L_{1},t \right)  =0.
\end{equation*}

The above equations are sufficient to determine  $\mathbf{A_{r}}$ and $\mathbf{A_{t}}$ from $\mathbf{A_{i}}$.

A similar set of equations are obtained for $\mathbf{E_{t}}$, the transmitted $\mathbf{E_{tt}}$ and reflected  $\mathbf{E_{tr}}$ fields at  $y=L_{2}$.  The matching conditions for the angular frequencies at this interface are:
\begin{equation*}
\omega_{t}(L_{2}, t)= \omega_{tt}(L_{2}, t) =\omega_{tr}(L_{2}, t),
\end{equation*}
with the corresponding conditions for the fields:
\begin{equation*}
\mathbf{E_{t}} \left( L_{2},t \right) + \mathbf{E_{tr}} \left( L_{2},t \right) -
\mathbf{E_{tt}}\left( L_{2},t \right)  =0,
\end{equation*}
\begin{equation*}
\mathbf{B_{t}} \left( L_{2},t \right) + \mathbf{B_{tr}} \left( L_{2},t \right) -
\mathbf{B_{tt}} \left( L_{2},t \right)  =0.
\end{equation*}

Taking $\mathbf{A_{t}}$ as the incident data, the above equations yield $\mathbf{A_{tt}}$ and $\mathbf{A_{tr}}$.

Finally, when the wave $\mathbf{E_{tr}}$ reaches $y=L_{1}$ a transmitted field $\mathbf{E_{trt}}$ that goes back to the initial media $n_1$ is produced.  The reflected wave is order $\gamma^{2}$ and thus, it is discarded.

The resulting matching condition for the angular frequency is given by:
\begin{equation*}
\omega_{tr}(L_{1}, t)= \omega_{trt}(L_{1}, t),
\end{equation*}

whereas the matching condition for the field is:
\begin{equation*}
\mathbf{E_{tr}} \left( L_{1},t \right) -\mathbf{E_{trt}} \left( L_{1},t \right)  =0,
\end{equation*}
and the last vector potential $\mathbf{A_{trt}}$ is obtained.

The final set of vector potentials that satisfy the angular frequency and amplitude matching conditions is given by
\small
\begin{equation}
\mathbf{A_{{i}}}(y,t)=A_{{0}}\ \mathrm {exp}{\left[\mathrm{i} \left({ \mathrm{k} y  -\omega_{0} t } \right)  \right]} \ \mathbf{\hat{e}_{z}},
\label{eq:SD1}
\end{equation}

\begin{equation}
\begin{aligned}
\mathbf{A_{t}}(y,t)= & \ A_{0} \left[ 1 + \frac{\gamma}{4} \cos(\Omega t) \right] \\ & \times \mathrm{exp}{\left( \mathrm{i} \left \{ \mathrm{k} y - \omega_{0} t - \frac{\omega_{0} \gamma}{\Omega} \sin\left[\frac{\Omega \left( y - L_{1} \right)}{2 v_{0}} \right] \right. \right.} \\ & \qquad \qquad  {\Bigg. \left.  \times \cos\left[ \Omega \left( t - \frac{y - L_{1}}{2 v_{0}} \right)\right] \right \}  \Bigg)} \ \mathbf{\hat{e}_{z}},
\end{aligned}
\end{equation}


\begin{equation}
\begin{aligned}
\mathbf{A_{tt}}(y,t) = & \  A_{0} \ \mathrm{exp}{\Bigg( \mathrm{i} \left \{ {\mathrm{k} y -\omega_{0}t} - \frac{\omega_{0} \gamma}{\Omega} \sin \left(\frac{\Omega L}{2 v_{0}} \right)  \right. \Bigg.}   \\ & \qquad \qquad {\Bigg. \left.  \times \cos\left[ \Omega \left( t - \frac{ L - 2 L_{2} + 2 y}{2 v_{0}} \right)\right] \right \} \Bigg)} \
\mathbf{\hat{e}_{z}},
\end{aligned}
\end{equation}


\begin{equation}
\begin{aligned}
\mathbf{A_{tr}}(y,t) = & - \frac{\gamma}{4} \ A_{0}  \  \cos\left[ \Omega
\left( t + \frac{ y - L_{2}}{v_{0}} \right)\right] \\ & \times \mathrm{exp}{\Bigg( -\mathrm{i} \left \{ \mathrm{k} y + \omega_{0} t - 2 \mathrm{k} L_{2}  \right. \Bigg.} \\  &
\qquad \quad + { \frac{\omega_{0} \gamma}{\Omega} \sin\left[\frac{\Omega
	\left( 2 L_{2} - L_{1} - y   \right)}{2 v_{0}}\right]  } \\  &
\qquad \quad \times  { \Bigg. \left. { \cos\left[\Omega \left( t - \frac{  2 L_{2} - L_{1} - y  }{2 v_{0}}  \right) \right] } \right \} \Bigg) } \ \mathbf{\hat{e}_{z}},
\end{aligned}
\end{equation}


\begin{equation}
\begin{aligned}
\mathbf{A_{r}}(y,t) = & \  \frac{\gamma}{4} \ A_{0} \  \cos\left[\Omega
\left( t - \frac{L_{1} - y}{v_{0}} \right)\right] \\ & \times \mathrm{exp}{\left[ - \mathrm{i} \left(  
	\mathrm{k} y  + \omega_{0} t - 2\mathrm{k} L_{1}  \right) \right] } \
\mathbf{\hat{e}_{z}},
\end{aligned} 
\label{eq:SD5}
\end{equation}


\begin{equation}
\begin{aligned}
\mathbf{A_{trt}}(y,t) = & - \frac{\gamma}{4} \
A_{0} \  \cos\left[ \Omega \left( t - \frac{L_{2} - y}{v_{0}} \right)\right] \\  & \times \mathrm{exp}{\Bigg(  -\mathrm{i}
	\left \{ \mathrm{k} y  + \omega_{0} t  - 2 \mathrm{k} L_{2}  
	\right. \Bigg.}  \\ & \quad \qquad +{\frac{\omega_{0} \gamma}{\Omega} \sin
	\left( \frac{\Omega L}{v_{0}}  \right)} \\ &  \qquad \quad
{\Bigg. \left. \times \cos \left[\Omega\left( t - \frac{ L_{2} - y}{v_0} \right) \right] \right \} \Bigg)}
\ \mathbf{\hat{e}_{z}}.
\end{aligned}
\label{eq:SD6}
\end{equation}


\normalsize

Note that  (\ref{eq:SD1})-(\ref{eq:SD6}) converge to the static solutions (\ref{eq:E1})-(\ref{eq:E3}) when the limit $\Omega \to 0$ is taken.
Likewise, $\mathbf{A_{i}}+\mathbf{A_{r}}+\mathbf{A_{trt}}$ coincides exactly with $\mathbf{A_{1}}$ in (\ref{eq:A1+}) and (\ref{eq:A1-}), as expected. Similarly, $\mathbf{A_{t}}+\mathbf{A_{tr}}$ coincides with $\mathbf{A_{2}}$ in  (\ref{eq:A2+}) and (\ref{eq:A2-}). Finally, $\mathbf{A_{tt}}$  is equal to $\mathbf{A_{3}}$ given in (\ref{eq:A3+}).

It is important to note that (\ref{eq:SD1})-(\ref{eq:SD6}) cannot be obtained using the usual Fresnel's 
equations. Indeed, Fresnel's equations provide a relation between the amplitude of the electric and magnetic fields at both interfaces ($y=L_{1}$ and $y=L_{2}$ in our cases). 

In the static case, these let resolve the problem 
because the amplitudes of the fields are constants and their phases are determined for wavenumber of the medium 
and the angular frequency of the incident wave, $\omega_{0}$ (see (\ref{eq:CE_A1})-(\ref{eq:CE_A3}) and 
(\ref{eq:Ez}),(\ref{eq:Bx})).

However, in the dynamic case, the amplitudes and phases of the fields satisfy a coupled partial differential equations system and the matching conditions give relations between the boundaries values of the amplitudes in both interfaces.   Indeed, in this case, both the wavenumbers and the angular frequencies of the fields depend on the spatial coordinate and the time.


\section{Computing the RAM fundamental}
\label{Appendix C}
\qquad In this appendix we compute the $\mathrm{RAM_{F}}$. 

In general, given  a signal emerging of the EOM, we can define $\mathrm{RAM}$ as
\begin{equation}
\mathrm{RAM} = \mathrm{\frac{AC}{DC}},
\label{eq:RAM}
\end{equation}
where $\mathrm{AC}$ is the alternating component and $\mathrm{DC}$ the continuous component of the photocurrent generated by the  emerging beam from EOM.

 $\mathrm{DC}$ is the first  coefficient in the Fourier expansion,
\begin{equation}
\mathrm{DC} = \frac{\Omega}{2 \pi}\,\int _{-\frac {\pi }{\Omega}}^{{\frac {\pi }{\Omega}}} \ i_{c}(\tau)  \ {\mathrm{d} \tau},
\label{eq:DC}
\end{equation}
where $i_{c}$ represents the photocurrent.
The alternating component $\mathrm{AC}$ is the magnitude of the  demodulation components $\mathcal{I}$ and $\mathcal{Q}$ of $i_{c}$, given by
\begin{equation}
\mathrm{AC} = \sqrt{\mathcal{I}^{2}+\mathcal{Q}^{2}},
\label{eq:AC}
\end{equation}
where
\begin{equation}
\mathcal{I} = \frac{\Omega}{\pi} \int _{-{\frac {\pi }{\Omega}}}^{{\frac {\pi }{\Omega}}} i_{c}(\tau) \ \cos
\left( \Omega\,\tau \right) {\mathrm{d} \tau},
\label{eq:inphase}
\end{equation}

\begin{equation}
\mathcal{Q} = \frac{\Omega}{\pi} \int _{-{\frac {\pi }{\Omega}}}^{{\frac{\pi }{\Omega}}} i_{c}(\tau) \ \sin\left( \Omega\, \tau \right) {\mathrm{d} \tau}.
\label{eq:quadrature}
\end{equation}

Now we compute the $\mathrm{RAM}$ for an ideal EOM that is to say $\mathrm{RAM_{F}}$.

Replacing (\ref{eq:current}) in (\ref{eq:DC})-(\ref{eq:quadrature}) we obtain the DC and AC components. Finally, 
from  (\ref{eq:RAM}) we obtain:

\begin{equation}
\mathrm{RAM_{F}} = \frac{2 m \Omega}{\omega_{0}}.
\end{equation}

\begin{backmatter}
\bmsection{Acknowledgements} This research has been supported by the Faculty of Engineering of Instituto Universitario Aeron\'autico and by CONICET. 
The authors wish to thank the LIGO Scientific Collaboration for sharing relevant information concerning $\mathrm{RAM}$ in LIGO. 
\end{backmatter}

\bibliography{references}
\bibliographystyle{osajnl}
\bibliographyfullrefs{references}
\end{document}